\documentclass[%
 prl,reprint,
superscriptaddress,
showpacs,%preprintnumbers,
 amsmath,amssymb,
 aps,
]{revtex4-1}

\usepackage{graphicx}
\usepackage[space]{grffile}
\usepackage{latexsym}
\usepackage{textcomp}
\usepackage{longtable}
\usepackage{multirow,booktabs}
\usepackage{amsfonts,amsmath,amssymb}
\usepackage{natbib}
\usepackage{url}
\usepackage{hyperref}
\hypersetup{colorlinks=false,pdfborder={0 0 0}}
\newif\iflatexml\latexmlfalse
\usepackage[utf8]{inputenc}
\usepackage[english]{babel}
\usepackage{epstopdf}

\begin{document}

\title{%Revealing Dark Excitons Through Their Coherent Interactions
Revealing and Characterizing Dark Excitons Through Coherent Multidimensional Spectroscopy}

 \author{Jonathan O. Tollerud}
 \affiliation{Centre for Quantum and Optical Science, Swinburne University of Technology, Hawthorn, Victoria, Australia, 3122}
 \author{Steven T. Cundiff}
 \affiliation{Department of Physics, University of Michigan, Ann Arbor, Michigan 48109, USA}
 \author{Jeffrey A. Davis}
 \affiliation{Centre for Quantum and Optical Science, Swinburne University of Technology, Hawthorn, Victoria, Australia, 3122}

% You should use BibTeX and apsrev.bst for references
% Choosing a journal automatically selects the correct APS
% BibTeX style file (bst file), so only uncomment the line
% below if necessary.
\bibliographystyle{apsrev4-1}

\begin{abstract}
Dark excitons are of fundamental importance in a broad range of contexts, but are difficult to study using conventional optical spectroscopy due to their weak interaction with light. We show how coherent multidimensional spectroscopy can reveal and characterize dark states. Using this approach, we identify parity forbidden and spatially indirect excitons in InGaAs/GaAs quantum wells and determine details regarding lifetimes, homogeneous and inhomogeneous linewidths, broadening mechanisms and coupling strengths. The observations of coherent coupling between these states and bright excitons hint at a role for a multi-step process by which excitons in the barrier can relax into the quantum wells.%
\end{abstract}%

\maketitle

Optical spectroscopy provides convenient tools to characterize excitations that interact strongly with light (`bright' transitions); 
Linear techniques can be used to determine the energies and linewidths, while more recent ultrafast techniques can be used to understand interaction processes, dynamics and even many-body effects~\cite{SHAH1999,Chemla2001,Cundiff2012,Tollerud2014,Cundiff2014}.
Excitations that interact weakly or not at all with light (`dark' transitions) are, however, more difficult to characterize or even identify, yet they are important in many systems and relevant to a wide variety of phenomena. For example, radiative recombination rates of bright excitons (bound electron-hole pairs) in colloidal~\cite{Nirmal1995,Efros1996,Crooker2003} and self-assembled~\cite{Moody2011,Bayer2000a} quantum dots cannot be described without considering the role of spin-forbidden excitons. Similarly, this type of dark exciton must be considered to fully understand population transfer between some nanostructures \cite{Moody2011}. In molecular systems, symmetry-forbidden dark states frequently play a major role in the relaxation pathways, but remain difficult to fully characterize. For example, the energy landscape in carotenoid molecules and the presence and role of symmetry-forbidden dark states in the relaxation pathways has been the subject of much debate~\cite{Tavan1987,Cerullo2002,Wohlleben2004b,Ostroumov2013}. For organic photovoltaics the extraction of energy necessarily involves a charge-transfer state in which the electron and hole are spatially separated and thus interact weakly with light~\cite{Vandewal2014,Bakulin2012,Muntwiler2008,Zhu2009}. A similar charge transfer state is crucial for charge separation in photosynthesis while recent proposals have suggested a role for other symmetry-forbidden dark states~\cite{Romero2014,Ferretti2016}. Our ability to gain a detailed understanding of these states and their roles is hampered by their weak interaction with light. 

Semiconductor quantum wells (QWs) offer a unique opportunity to study different types of excitons because their states, transitions and interactions can be engineered. While QWs have typically been designed to maximize their interactions with light, motivated by device applications such as lasers and detectors \cite{Faist1994,Brown1991,Davis2012}, the ability to engineer excitons that interact weakly with light has also been important in, for example, measurements of exciton transport~\cite{Snoke2002} and efforts towards Bose-Einstein condensation of excitons~\cite{High2012,Larionov2001,Butov2002}. 
Various QW excitons with reduced dipole oscillator strength can exist: spatially indirect excitons~\cite{Marzin1985,Moran1998,Danan1987,Wilson1988} can arise due to type-II band alignment or in coupled QWs. In this case the oscillator strength is highly dependent on the electron hole overlap and the resultant excitons are analogous to the CT states in molecular systems~\cite{Vandewal2014,Bakulin2012,Muntwiler2008,Zhu2009}. Parity forbidden excitons, which involve electrons and holes with opposite parity, typically have transition dipoles that are much weaker, becoming weakly allowed with %the strict {forbiddenness} lifted by 
any asymmetry in the QW potential profiles~\cite{Wang1995,Yu2011,Shen1986,Chen1999,Kwok1992,Pan1988,Shen1994}. These excitons are equivalent to the symmetry forbidden transitions in molecular systems. Finally, spin forbidden excitons that have total angular momentum 0 or $\pm 2 \hbar$ typically have zero transition dipole moment, although phonon-assisted dipole transitions can become possible~\cite{Efros1996,Crooker2003} and quadrupole transitions are allowed.
%}.  

Various methods have been used to identify these states. For example, selection rules can be relaxed by applying a magnetic field or some structural change that can mix spin states. Alternatively, coupling to phonons can help overcome strict selection rules~\cite{Efros1996,Crooker2003}. Under the right circumstances, intrinsic optical signals can even be detected for each of the excitons described above.
However, while these approaches can identify dark states, the detailed characterization that optical spectroscopy provides for bright states is still missing.

Here we show that coherent multidimensional spectroscopy (CMDS)~\cite{Cho2008,Cundiff2012,Nardin2016a,Cundiff2014,Tollerud2014} can be used to reveal and characterize dark states. CMDS is a powerful extension of four-wave mixing (FWM) spectroscopy whereby signals are spread out along two or more frequency axes so they can be separated based on the quantum mechanical pathways that generate them: pathways that involve interactions between spectrally distinct transitions appear as cross-peaks (CPs) while pathways that involve only a single transition or spectrally degenerate transitions appear as diagonal-peaks (DPs).
In this letter, we use CMDS to reveal and characterize parity forbidden and spatially indirect excitons that are strongly coupled to bright states and thereby produce strong, easily identifiable CPs. We exploit the fact that the amplitude of a DP for a transition `$i$' (with dipole moment $\mu[i]$) is $\propto \mu[i]^4$, while the amplitude for a CP involving interactions between a dark state (d) and a bright state (b) is $\propto J[b,d]\times \mu[b]^2\times \mu[d]^2$, where $J[b,d]$ is a phenomenological coupling factor ranging from 0 (uncoupled) to 1 (strongly coupled). If dark and bright states are strongly coupled %($J[b,d]\approx 1$) 
and $\mu[b] \gg \mu[d]$ then the CP amplitude will 
%be much larger than the dark state DP amplitude, 
dwarf that of the dark state DP, allowing CPs involving the dark state to be detected even when the dark state DP cannot. It is even possible to excite pathways that drive measureable emission from the dark states. Such peaks can be identified easily in 2D spectra, and new details become accessible in 3D spectra. 
Using this approach, we can directly probe these states with an all optical experiment.
Moreover, CMDS can reveal useful details about the dark states that other techniques cannot: spectral linewidths, population lifetimes, coherent dynamics, broadening mechanisms and relative coupling strengths.

We performed the CMDS experiments using a diffraction based pulse-shaper similar to the setup originally developed by Nelson \textit{et al.}~\cite{Turner2011}. 
Experimental details can be found in the Supplemental Material (SM)~\cite{SuppMatt}, and a thorough description of our experimental apparatus can be found elsewhere~\cite{Tollerud2014}.
%Experimental details can be found in the supplemental material (SM), and a thorough description of our experimental apparatus can be found elsewhere~\cite{Tollerud2014}.

\begin{figure}[t]
\begin{center}
\includegraphics[width=1\columnwidth]{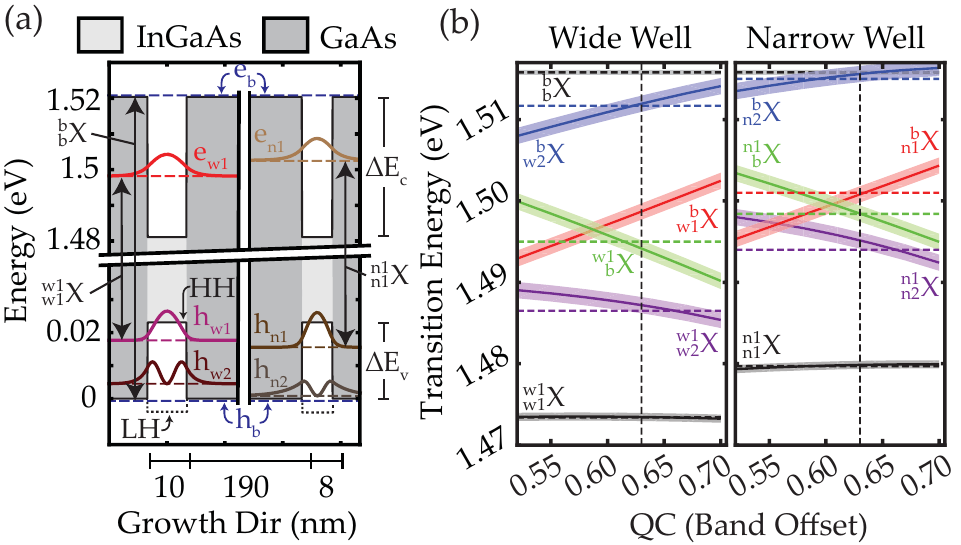}
\caption{(a) Diagram of the conduction band and valence band (HH \& LH) potentials, and typical calculated steady state wavefunctions for QW electrons and holes. (b) Calculated exciton transition energy for the various detected states as a function of QC, which coincide with the measured transition energies (horizontal dashed lines) at QC$\approx$0.63.  \label{Fig1}%
}
\end{center}
\end{figure}

\newcommand{\XX}[2]{$^{#1}_{#2} \textup{X}$}

The MBE grown sample consists of two independent In$_{0.05}$Ga$_{0.95}$As/GaAs QWs, 8\,nm and 10\,nm wide, separated by a 190\,nm barrier and a 5 period AlAs/GaAs superlattice to suppress any carrier migration between the wells. %
%In In$_x$Ga$_{1-x}$As/GaAs QWs with x\textless0.2, strain
Strain in the In$_{0.05}$Ga$_{0.95}$As layer shifts the light hole (LH) valence band below the GaAs valence band, leading to type-II LH excitons, while the HH excitons remain type-I. This coexistence of stable type-I and type-II excitons presents an intriguing case for exploring interactions between bright and dark states and by exploiting their coupling, spectroscopic tools that are normally reserved for bright states can be applied to reveal details of dark states.

Calculations of the steady-state wavefunctions (Fig.~\ref{Fig1} (a)) reveal that there are a total of one bound conduction band state in each QW ($e_{w1}$ and $e_{n2}$) and two bound HH valence band states in each well ($h_{w1}$ and $h_{w2}$ for the wide well, $h_{n1}$ and $h_{n2}$ for the narrow well). 
%In addition, there are electron (hole) states in the conduction band ($E_{Bar}$) (valence band ($H_{Bar}$)) of the GaAs barriers. 
These five~\footnote{There are six valence band states, but the GaAs HH and LH bands are degenerate and experimentally indistinguishable.} valence band states and three conduction band states combine to form a total of three direct, parity allowed exciton transitions: excitons localized in the wide well ($^{w1}_{w1} \textup{X}$), narrow well ($^{n1}_{n1} \textup{X}$), and barrier ($^{b}_{b}\textup{X}$). 
There are also two types of nominally dark exciton transitions: spatially direct, but parity forbidden transitions ($^{w1}_{w2} X$, $^{n1}_{n2} \textup{X}$); and spatially indirect excitons involving either a hole in the barrier and an electron in a QW ($^{w1}_{b} X$, $^{n1}_{b} \textup{X}$)or a hole in a QW and an electron in the barrier ($^{b}_{w1} \textup{X}$, $^{b}_{n1} \textup{X}$, $^{b}_{w2} \textup{X}$, $^{b}_{n2} \textup{X}$).

%The transition energies for these nominally dark transitions can be calculated based on the energy separation of the hole and electron states less the exciton binding energy. 
%The accuracy of the calculated transition energy relies on the accuracy of the material parameters, most of which are precisely defined for GaAs, but less so for strained In$_x$Ga$_{1-x}$As.

The accuracy of calculated transition energies for these nominally dark transitions -- which can be calculated based on the energy separation of the hole and electron states less the exciton binding energy -- relies on the accuracy of the material parameters, most of which are precisely defined for GaAs, but less so for strained In$_x$Ga$_{1-x}$As.
%For example, the In$_{0.05}$Ga$_{0.95}$As band offset (QC) has been reported to be from 0.46 to 0.75~\cite{Chi1995,Arent1989,Moore1990,Joyce1991}. 

%\begin{figure}[h!]
%\begin{center}
%\includegraphics[width=1\columnwidth]{Fig2.pdf}
%\caption{(a) Typical calculated steady state wavefunctions for electrons and holes confined in the QWs. (b) Calculated exciton transition energy for the dark %states as a functon of QC, which coincide with the measured transition energies (dashed lines) at QC$\approx$0.63. \label{WF_Calc}%
%}
%\end{center}
%\end{figure}

This uncertainty thus impedes our ability to use calculated transition energies to identify the experimentally detected signals.
To overcome this challenge we calculate transition energies using material constants across the reasonable parts of the parameter-space. The calculations were constrained so that the calculated transition energies for $^{w1}_{w1} \textup{X}$ and $^{n1}_{n1} \textup{X}$ match the easily identifiable experimental peaks. Calculated transition energies are presented in Fig.~\ref{Fig1} (b) as a function of the band-offset (QC), which is the least well defined and the key parameter for determining the spectral ordering of the transitions. QC is defined here as the fraction of the band gap difference occurring in the conduction band ($\Delta E_c / (\Delta E_c + \Delta E_v $). In order to fix the energies of $^{w1}_{w1} \textup{X}$ and $^{n1}_{n1} \textup{X}$ the indium content was varied linearly from 4.9\% at QC = 0.5 to  5.14\% at QC = 0.7.

The calculated and measured transition energies match at QC$\approx$0.63, which falls in the middle of the range of QC values previously reported~\cite{Chi1995,Arent1989,Moore1990,Joyce1991}. A detailed explanation of the parameters used and how they were varied to derive this plot is included in the SM~\cite{SuppMatt}.

%\begin{figure}[h!]
%\begin{center}
%\includegraphics[width=1\columnwidth]{Fig3.pdf}
%\caption{An absolute value rephasing 1Q 2D spectrum with a logarithmic colour scale. A number of CPs are observed due to the strong coupling of the bright and dark states.\label{Fig:CH4:PLE}%
%}
%\end{center}
%\end{figure}

\begin{figure}[t]
\begin{center}
\includegraphics[width=1\columnwidth]{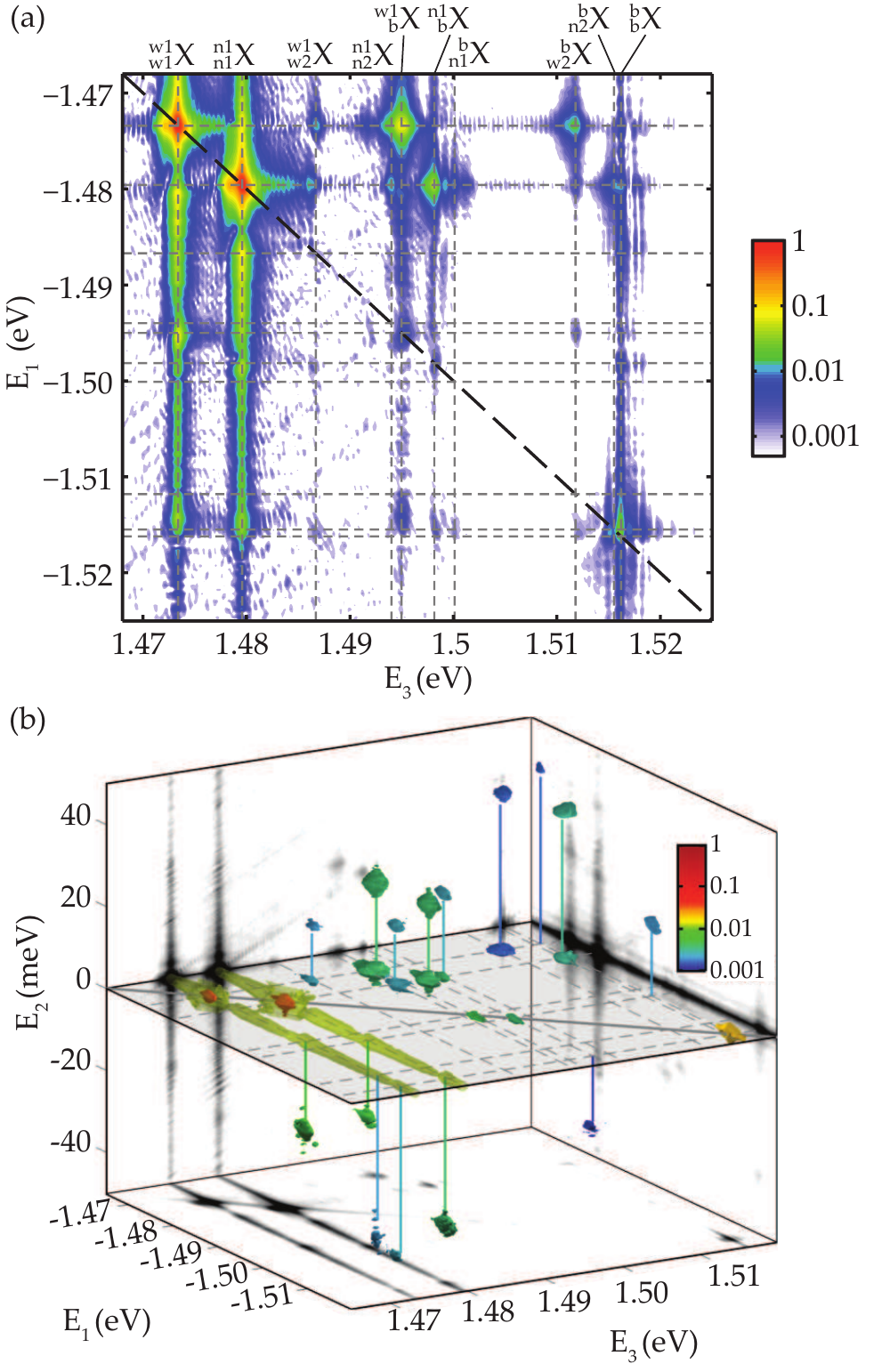}
\caption{(a) Absolute value rephasing 1Q 2D spectrum with a logarithmic colour scale. A number of CPs are observed due to the strong coupling of the bright and dark states. (b) An absolute value rephasing 3D spectrum. The amplitude of the detected peaks span several orders of magnitude, so for clarity each region is plotted at the most appropiate isosurface level.\label{Fig2}%
}
\end{center}
\end{figure}

A 2D spectrum acquired with three identical $\sim$40\,fs pulses in the rephasing pulse ordering is presented in Fig.~\ref{Fig2} (a). It is first important to note the logarithmic colour scale, spanning three orders of magnitude, which reveals several CPs even in the presence of much stronger bright state DPs. There are three DPs corresponding to $^{w1}_{w1} \textup{X}$, $^{n1}_{n1} \textup{X}$, and $^{b}_{b} \textup{X}$, two much lower amplitude DPs corresponding to $^{w1}_{b} \textup{X}$ and $^{n1}_{b} \textup{X}$, and at least 18 CPs corresponding to interactions between $^{w1}_{w1} \textup{X}$, $^{n1}_{n1} \textup{X}$, $^{b}_{b} \textup{X}$ and the parity forbidden and spatially indirect states. The spectra in Fig.~\ref{Fig2} are plotted using the convention that $E_1$ is negative due to the conjugated contribution from the first excitation field ($k_1$ in this case)~\cite{Cundiff2012}.

The below-diagonal CPs with emission at the QW energies overlap continua that arise due to coupling of $^{w1}_{w1} \textup{X}$ and $^{n1}_{n1} \textup{X}$ to unbound free-carriers, making it difficult to discern excitonic peaks. 
Several clear above-diagonal CPs peaks can be observed at [$E_3$, $E_1$] energies indicating coupling of the weakly allowed states to $^{w1}_{w1} \textup{X}$ and $^{n1}_{n1} \textup{X}$. 
%The three strongest CPs appear at $[E_3$, $E_1]$ = [1.495\,eV, -1.473\,eV], [1.498\,eV, -1.479\,eV] and [1.512\,eV, -1.473\,eV] which we attribute to [\XX{w1}{b}, \XX{w1}{w1}], [\XX{n1}{b}, \XX{n1}{n1}], and [\XX{b}{w2}, \XX{w1}{w1}], respectively. These pairs of transitions are also observed in photoluminescence excitation spectroscopy (see SM). 
The three strongest CPs appear at $[E_3$(eV), $E_1$(eV)$]$ = [1.495, -1.473], [1.498, -1.479] and [1.512, -1.473] which we attribute to [\XX{w1}{b}, \XX{w1}{w1}], [\XX{n1}{b}, \XX{n1}{n1}], and [\XX{b}{w2}, \XX{w1}{w1}], respectively. These pairs of transitions are also observed in photoluminescence excitation spectroscopy (see SM~\cite{SuppMatt}). 
%CPs are also resolved at [1.486\,eV, -1.473\,eV], [1.494\,eV, -1.479\,eV], [1.500\,eV, -1.479\,eV] and [1.515\,eV, -1.479\,eV] which we attribute to [\XX{w1}{w2}, \XX{w1}{w1}], [\XX{n1}{n2}, \XX{n1}{n1}], [\XX{b}{n1}, \XX{n1}{n1}], and [\XX{b}{n2}, \XX{n1}{n1}], respectively.
CPs are also resolved at [1.486, -1.473], [1.494, -1.479], [1.500, -1.479] and [1.515, -1.479] which we attribute to [\XX{w1}{w2}, \XX{w1}{w1}], [\XX{n1}{n2}, \XX{n1}{n1}], [\XX{b}{n1}, \XX{n1}{n1}], and [\XX{b}{n2}, \XX{n1}{n1}], respectively.
All of the electron and hole states involved in each of the detected CPs are localized in the same well, thus confirming our state attributions~\footnote{The [1.4855\,eV, -1.479\,eV] CP is attributed to coupling of $^{n1}_{n1} \textup{X}$ to 0-momentum free-carriers in the narrow QW.}. CPs involving the same pairs of transitions evident in the 1Q spectrum also appear in 0Q and 2Q 2D spectra (see SM~\cite{SuppMatt}), confirming that the parity forbidden and spatially indirect excitons are strongly and coherently coupled to the bright QW excitons.

We recorded a 3D spectrum to isolate and quantify the different CP contributions by Fourier transforming the data as a function of $t_2$. 
In this type of 3D spectrum, pathways involving a coherent superposition during $t_2$ generate peaks centered at a $E_2=\pm\Delta E_{CS}$ where $\Delta E_{CS}$ is the difference between the transition energies of the two states in superposition. A positive (negative) sign indicates that the phase evolution will be in the same (opposite) direction as the third pulse and hence that the energy will be added (subtracted) to the pulse 3 energy. In contrast, pathways that involve a population in $t_2$ (e.g. ground state bleach, excited state absorption) are centered at $E_2 = 0$.
%{generate peaks centered at a $E_2=\pm\Delta E_{CS}$ where $\Delta E_{CS}$ is the difference between the transition energies of the two states in superposition. A positive (negative) sign indicates that the phase evolution will be in the same (opposite) direction as the third pulse and hence that the energy will be added (subtracted) to the pulse 3 energy. In contrast, pathways that involve a population in $t_2$ (e.g. ground state bleach, excited state absorption) are centered at $E_2 = 0$. }
%In this type of 3D spectrum, pathways involving a coherent superposition during $t_2$ are shifted away from $E_2 = 0$, whereas pathways that involve a population in t$_2$ (e.g. ground state bleach, excited state absorption) are centered at $E_2 = 0$. 
%We also collected a 3D spectrum to better isolate the different signal pathways and extract useful information beyond simply detecting the dark states, (e.g. quantifying the coupling strengths and population lifetimes).
%In order to quantify the coupling strengths, determine peak shapes and population lifetimes we need to better isolate the different signal pathways, which we achieve with a 3D spectrum~\cite{Davis2011a,Hall2013,Cundiff2014,Tollerud2014}. 
When we go from 2D to 3D, overlapping pathways are unravelled revealing additional information, just as it did when we went from 1D to 2D~\cite{Li2013}. For example, from the 3D spectrum we are able to determine coupling strengths, peakshapes and lifetimes.

%\begin{figure}[h!]
%\begin{center}
%\includegraphics[width=1\columnwidth]{Fig4.png}
%\caption{An absolute value rephasing 3D spectrum. The amplitude of the detected peaks span several orders of magnitude, so each region is plotted at an isosurface level. \label{Fig:CH4:30nm_3DSpectrum}%
%}
%\end{center}
%\end{figure}

The 3D spectrum shown in Fig.~\ref{Fig2} (b) has different regions rendered on different isosurface levels in order to show all the pertinent data, with amplitudes spanning more than three orders of magnitude, in a single figure. We are able to isolate a total of 33 separate peaks, 5 DPs, 13 population CPs and 15 coherence CPs. 
A detailed analysis of this 3D spectrum is included in the SM~\cite{SuppMatt}.

To determine the coupling strengths the peak amplitudes first need to be normalized for the spectral weight of the pulses at each interaction energy (see SM for details~\cite{SuppMatt}) and the relative dipole moments determined from DP amplitudes, where possible.  %Among the `dark' excitons diagonal peaks can be seen only for $\beta X_1$ and $\beta X_2$, making quantitative assessment of coupling strengths involving other dark states difficult. 
Relative coupling strengths are then determined based on the relative amplitudes of the CP signals S$_{m,n}$ which are normalized according to some reference amplitude $S_0$: 

\begin{equation}
	 \textup{J}[m,n] = \frac{S_{m,n}}{S_0} \frac{1}{\widetilde{\mu}[m]^2 \widetilde{\mu}[n]^2} 
\end{equation}
where $\textup{J}[m,n]$ is a phenomenological coupling constant intended to characterize the strength of the interactions between the two transitions.

%\begin{table} 
%    \begin{tabular}{ c | c c c c c }
%        E$_1$/E$_3$ & $X_W$ & $X_N$ & $\beta X_W$ & $\beta X_N$ & $X_{Bar}$ \\ 
%        \hline
%        $X_W$ & - &  &  0.44 &  &  0.014 \\ 
%        $X_N$ &  & - &  & 0.37 & 0.04 \\ 
%        $\beta X_W$ & 0.39 &  & - &  &  \\ 
%        $\beta X_N$ &  & 0.37 &  & - & 0.25 \\ 
%        $X_{Bar}$ & 0.03 &  &  & 0.22 & - \\ 
%    \end{tabular} 
%    \caption{Relative coupling strengths (J$_{m,n}$) of coherence CPs for which both of the associated DPs are detected.\label{TabCouple}} 
%\end{table}

%\newcommand{\JJ}[4]{$\textup{J}[ ^{#1}_{#2}\textup{X}, ^{#3}_{#4}\textup{X}]$ }
\newcommand{\JJnoJ}[4]{$[ ^{#1}_{#2}\textup{X},\, ^{#3}_{#4}\textup{X}]$}
\newcommand{\JJ}[4]{$\textup{J}$\JJnoJ{#1}{#2}{#3}{#4}}
\newcommand{\CP}[4]{$\textup{CP}$\JJnoJ{#1}{#2}{#3}{#4}}

\begin{table} 
    \begin{tabular}{c | c c | c | c c }
%        $X_W \leftrightarrow \beta X_W$ & $X_N \leftrightarrow \beta X_N$ & $X_{Bar} \leftrightarrow X_W$ & $X_{Bar} \leftrightarrow X_N$ & $X_{Bar} \leftrightarrow \beta X_N$ \\ 
%        $X_W$, $\beta X_W$ & $X_N$, $\beta X_N$ & $X_{Bar}$, $X_W$ & $X_{Bar}$, $X_N$ & $X_{Bar}$, $\beta X_N$ \\ 
        %&$^{w1}_{w1} \textup{X}$--$^{w1}_{b} \textup{X}$ & $^{n1}_{n1} \textup{X}$--$^{w1}_{b} \textup{X}$ &$^{b}_{b} \textup{X}$-- $^{n1}_{b} \textup{X}$ & $^{b}_{b} \textup{X}$--$^{w1}_{w1} \textup{X}$ & $^{b}_{b} \textup{X}$--$^{n1}_{n1} \textup{X}$\\  \hline
        &\JJnoJ{w1}{b}{w1}{w1}& \JJnoJ{n1}{b}{n1}{n1}& \JJnoJ{n1}{b}{b}{b}& \JJnoJ{w1}{w1}{b}{b}& \JJnoJ{n1}{n1}{b}{b}\\  \hline
	%0.44 (0.39) & 0.37 (0.37) & 0.25 (0.22) & 0.014 (0.03) & 0.04 \\
	AD & 0.44 & 0.37 & 0.25 & 0.014 & 0.04  \\
	BD & 0.39 & 0.37 & 0.22 & 0.03 & - \\ \hline
    \end{tabular} 
    \caption{$\textup{J}[m,n]$ (relative coupling strengths) for above diagonal (AD) and below diagonal (BD) coherence CPs for which both of the associated DPs are detected.\label{TabCouple}} 
\end{table}

%{The J values extracted from the 3D spectrum are shown in Table~\ref{TabCouple}.  \JJ{w1}{b}{w1}{w1}, \JJ{n1}{b}{n1}{n1}, and \JJ{n1}{b}{b}{b} are roughly an order of magnitude larger than \JJ{w1}{w1}{b}{b} and \JJ{n1}{n1}{b}{b}. This enhancement can be explained by increased spatial overlap and by a the shared electron state (for \JJ{w1}{b}{w1}{w1}, \JJ{n1}{b}{n1}{n1}) or hole state (for \JJ{n1}{b}{b}{b}).}
The J values extracted from the 3D spectrum are shown in Table~\ref{TabCouple}. The $\approx 10\times$ enhancement of coupling between spatially indirect and bright excitons (\JJ{w1}{b}{w1}{w1}, \JJ{n1}{b}{n1}{n1}, and \JJ{n1}{b}{b}{b}) relative to the coupling between QW and barrier excitons (\JJ{w1}{w1}{b}{b} and \JJ{n1}{n1}{b}{b}) can be explained by increased spatial overlap and by the shared electron state (\JJ{w1}{b}{w1}{w1}, \JJ{n1}{b}{n1}{n1}) or hole state (\JJ{n1}{b}{b}{b}).

\newcommand{\TMu}[2]{$\tilde{\mu}[ ^{#1}_{#2}\textup{X}]$ }

While we do not observe any other DPs, the CP amplitudes still provide an indication of the coupling strength. 
The normalized amplitude of \CP{b}{w2}{w1}{w1} and \CP{b}{n2}{n1}{n1} are similar to that of \CP{w1}{b}{w1}{w1} and \CP{n1}{b}{n1}{n1}, and
%The normalized $^{b}_{w2} \textup{X}$--$^{w1}_{w1} \textup{X}$ and $^{b}_{n2} \textup{X}$--$^{n1}_{n1} \textup{X}$ CPs are similar in amplitude to the $^{w1}_{b} \textup{X}$--$^{w1}_{w1} \textup{X}$ and $^{n1}_{b} \textup{X}$--$^{n1}_{n1} \textup{X}$ CPs.
it can be assumed based on the absence of $^{b}_{w2} \textup{X}$ or $^{b}_{n2} \textup{X}$ DPs that \TMu{b}{w2} and \TMu{b}{n2} are no larger than \TMu{w1}{b} and \TMu{n1}{b}~\footnote{The lower pulse amplitude at the energy of $^{b}_{w2} \textup{X}$ and $^{b}_{n2} \textup{X}$ means that the transition dipole strength could be similar to that of \XX{w1}{b} and \XX{n1}{b} but simply not visible.}. It is thus reasonable to conclude that \JJ{b}{w2}{w1}{w1} and \JJ{b}{n2}{n1}{n1} are comparable to or larger than \JJ{w1}{b}{w1}{w1} and \JJ{n1}{b}{n1}{n1}. 
Coupling between the parity forbidden excitons and the bright QW excitons is expected to be larger due to high spatial overlap and a shared electron state, while the parity forbidden oscillator strengths are expected to be lower. The corresponding % parity forbidden ($^{w1}_{w2} \textup{X}$ and $^{n1}_{n2} \textup{X}$) 
CPs are an order of magnitude lower than the spatially indirect CPs, which indicates that the transition dipole moments are more than an order of magnitude lower than for the spatially indirect states, which in turn are almost an order of magnitude lower than for the bright states.%. $\alpha X$ are, however, expected to be the darkest excitons with $\tilde{\mu}_{\delta X}$ at least two orders of magnitude weaker than $\tilde{\mu}_{X}$, suggesting $J_{\alpha X, X}$ may be significantly larger than $J_{\beta X, X}$ and $J_{\delta X, X}$.

In contrast to previous work -- which determined that exciton capture is largely ambipolar \cite{Blom1993} and may be mediated by coherent coupling of barrier and QW excitons~\cite{Reynolds1991,Reynolds1993} -- our observations suggest that ambipolar capture due to direct barrier-QW coherent coupling may compete with a two-step process where $^{b}_{b} \textup{X}$ is coherently coupled to the spatially indirect states, which in turn are strongly coupled to the direct QW excitons.

In addition to identifying these parity forbidden and spatially indirect excitons and determining coupling strengths, various details can also be extracted from the shapes of the 3D peaks and their projections. Information on the population dynamics can be obtained directly from the 3D spectrum due to the inverse proportionality of linewidths and lifetimes. The population lifetimes of $^{w1}_{w1} \textup{X}$, $^{n1}_{n1} \textup{X}$, $^{w1}_{b} \textup{X}$, and $^{n1}_{b} \textup{X}$ can be determined from the DPs centered at $E_2 = 0$, and are inversely proportional to the widths along the $E_2$ axis. This analysis yields population lifetimes of $\geq$800\,fs for  $^{w1}_{b} \textup{X}$ and $^{n1}_{b} \textup{X}$, which are comparable to our measurements of $^{w1}_{w1} \textup{X}$ and $^{n1}_{n1} \textup{X}$. These measurements, however, are limited by the delay range of the pulse shaper (see SM~\cite{SuppMatt}) and so actually represent a lower bound for the lifetimes~\footnote{The experimental limitations that restrict us to presenting lower bounds on lifetimes are specific to this implementation of CMDS; the delay range can be extended by using translation stages instead of a pulse shaper to control the pulse timings.}. A lower bound for the lifetimes of the other spatially indirect and parity forbidden excitons, for which we have not observed DPs, can be obtained from the coherence CPs. In this case the width of the CPs along $E_2$ is a convolution of the pure decoherence time of the coherent superposition and the lifetimes of the individual states in the superposition. The measured lower bound of $\sim$500\,fs, while not the long radiative lifetime expected for dark states, does indicate that each of the excitons persist well beyond pulse overlap and do not rapidly relax non-radiatively into other states (e.g. $^{w1}_{w1} \textup{X}$ or $^{n1}_{n1} \textup{X}$).

\begin{figure}[t]
\begin{center}
\includegraphics[width=1\columnwidth]{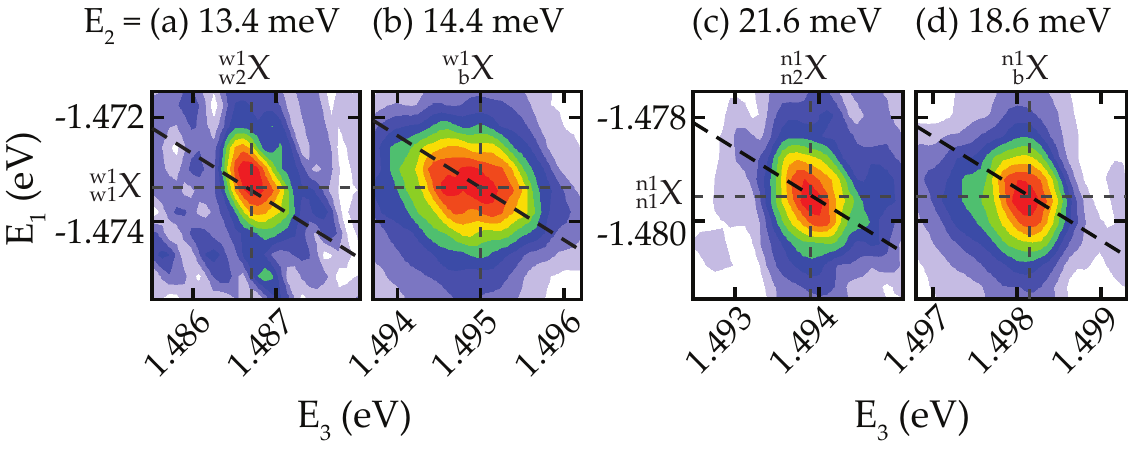}
\caption{Projections of several 3D CPs onto the $E_1$ vs $E_3$ plane, which reveal details about the spectral broadening of the dark states, as discussed in the text. \label{Fig5}%
}
\end{center}
\end{figure}

%It is well established in 2D spectroscopy that information about broadening mechanisms can be extracted from peak shapes
It is well established that information about broadening mechanisms can be extracted from 2D peak shapes~\cite{Bristow2011,Siemens2010,Tollerud2016PS}. For dark states the shape and tilt of the relevant CPs reveal details of inhomogeneous broadening, including whether there is any correlation with the broadening of the bright QW exciton state to which it is coupled~\cite{Davis2011a,Tollerud2014}. Figure~\ref{Fig5} shows projections of 
%the $^{w1}_{w2} \textup{X}$--$^{w1}_{w1} \textup{X}$, $^{w1}_{b} \textup{X}$--$^{w1}_{w1} \textup{X}$, $^{n1}_{n2} \textup{X}$--$^{n1}_{n1} \textup{X}$, and $^{n1}_{b} \textup{X}$--$^{n1}_{n1} \textup{X}$ 
several coherence CPs, which all exhibit a slight tilt towards the diagonal, indicating some correlation of the inhomogeneous broadening. The more pronounced tilt in (a) and (c) is expected given that $^{w1}_{w2} \textup{X}$ and $^{n1}_{n2} \textup{X}$ are contained entirely within the QW (and as such experience the same static disorder as $^{w1}_{w1} \textup{X}$ and $^{n1}_{n1} \textup{X}$, respectively), whereas the holes in $^{w1}_{b} \textup{X}$ and $^{n1}_{b} \textup{X}$ are not confined in the QW, (reducing the correlation of the inhomogeneous broadening). The steeper than 1:1 tilt of the CPs in (a) and (c) further indicates that 
%although the inhomogeneous broadening is correlated, 
the amount of broadening is reduced for $^{w1}_{w2} \textup{X}$ and $^{n1}_{n2} \textup{X}$ compared to $^{w1}_{w1} \textup{X}$ and $^{n1}_{n1} \textup{X}$. This in turn suggests that the energy of $h_{w1}$ and $h_{n1}$ are more sensitive to the prevailing disorder than $h_{w2}$ and $h_{n2}$.

In summary, we have demonstrated that multidimensional spectroscopy is an effective tool for studying dark excitons. We have observed parity forbidden and spatially indirect states in strained GaAs/InGaAs QWs by exploiting the fact that they are strongly coupled to bright states, which effectively amplifies the signal from these dark states and makes them more optically accessible. Further refinement of the technique will allow the detection of excitons with even weaker transition dipole moments, potentially including the spin forbidden excitons.  Beyond just detecting the states we have shown that we can leverage the power of CMDS to extract useful information, including relative coupling strengths, lifetimes, spectroscopic linewidths and broadening mechanisms. 
CMDS allows us to reveal and characterize the rich structure of dark states that is otherwise hidden and can open the door to previously unexplored physics in a variety of excitonic systems.

\begin{acknowledgments}
The authors thank G. Nardin and F. Morier-Genoud for providing us with the QW sample and for helpful conversations. JD and JT thank the Australian Research Council for funding support.
\end{acknowledgments}

\end{document}